\documentclass[aps,pra,preprint,longbibliography,superscriptaddress,showpacs,showkeys]{revtex4-1}
\usepackage{graphicx}
\usepackage{dcolumn}
\usepackage{bm}
\usepackage{color}
\usepackage{lmodern}
\usepackage[toc,page]{appendix}

\begin{document}

\author{F. Heyroth}
\affiliation{Interdisziplin\"{a}res Zentrum f\"{u}r Materialwissenschaften, Martin-Luther-Universit\"{a}t Halle-Wittenberg, D-06120 Halle, Germany}

\author{C. Hauser}
\affiliation{Institut f\"{u}r Physik, Martin-Luther-Universit\"{a}t Halle-Wittenberg, D-06120 Halle, Germany}

\author{P. Trempler}
\affiliation{Institut f\"{u}r Physik, Martin-Luther-Universit\"{a}t Halle-Wittenberg, D-06120 Halle, Germany}

\author{P. Geyer}
\affiliation{Institut f\"{u}r Physik, Martin-Luther-Universit\"{a}t Halle-Wittenberg, D-06120 Halle, Germany}

\author{F. Syrowatka}
\affiliation{Interdisziplin\"{a}res Zentrum f\"{u}r Materialwissenschaften, Martin-Luther-Universit\"{a}t Halle-Wittenberg, D-06120 Halle, Germany}

\author{R. Dreyer}
\affiliation{Institut f\"{u}r Physik, Martin-Luther-Universit\"{a}t Halle-Wittenberg, D-06120 Halle, Germany}

\author{S.G. Ebbinghaus}
\affiliation{Institut f\"{u}r Chemie, Martin Luther Universit\"{a}t Halle-Wittenberg, D-06120 Halle, Germany}

\author{G. Woltersdorf}
\affiliation{Institut f\"{u}r Physik, Martin-Luther-Universit\"{a}t Halle-Wittenberg, D-06120 Halle, Germany}

\author{G. Schmidt}
\email[Correspondence to G. Schmidt: ]{georg.schmidt@physik.uni-halle.de}
\affiliation{Institut f\"{u}r Physik, Martin-Luther-Universit\"{a}t Halle-Wittenberg, D-06120 Halle, Germany}
\affiliation{Interdisziplin\"{a}res Zentrum f\"{u}r Materialwissenschaften, Martin-Luther-Universit\"{a}t Halle-Wittenberg, D-06120 Halle, Germany}

\title{Monocrystalline free standing 3D yttrium iron garnet magnon nano resonators}

\begin{abstract}
Nano resonators in which mechanical vibrations and spin waves can be coupled are an intriguing concept that can be used in quantum information processing to transfer information between different states of excitation. Until now, the fabrication of free standing magnetic nanostructures which host long lived spin wave excitatons and may be suitable as mechanical resonators seemed elusive. We demonstrate the fabrication of free standing monocrystalline yttrium iron garnet (YIG) 3D nanoresonators with nearly ideal magnetic properties. The freestanding 3D structures are obtained using a complex lithography process including room temperature deposition and lift-off of amorphous YIG and subsequent crystallization by annealing. The crystallization nucleates from the substrate and propagates across the structure even around bends over distances of several micrometers to form e.g. monocrystalline resonators as shown by transmission electron microscopy. Spin wave excitations in individual nanostructures are imaged by time resolved scanning Kerr microscopy. The narrow linewidth of the magnetic excitations indicates a Gilbert damping constant of only $\alpha = 2.6 \times 10^{-4}$ rivalling the best values obtained for epitaxial YIG thin film material. The new fabrication process represents a leap forward in magnonics and magnon mechanics as it provides 3D YIG structures of unprecedented quality. At the same time it demonstrates a completely new route towards the fabrication of free standing crystalline nano structures which may be applicable also to other material systems.


\end{abstract}

\maketitle

\section{Introduction}

Nanomechanical oscillators are useful tools for quantum information processing. Over the past decade numerous groups have for example demonstrated the conversion of quantum information from the microwave to the optical regime by means of a micromechanical resonator\cite{Reed2017, Andrews2015, Lecocq2016, Ockeloen2016}\,. By coupling of electrical excitations in superconducting qubits to mechanical oscillators\cite{OConnell2010} even readout of quantum information has been demonstrated\cite{Lahaye2009, Chu2017}\,. The necessary interaction was often obtained by electric fields as in capacitive drum resonators. Another suitable mechanism for information transfer, however, can make use of the coupling of magnetic fields to spin wave modes in a magnon resonator. Indeed the coupling of a magnon mode in a macroscopic yttrium iron garnet (YIG) sphere to a single qubit has already been demonstrated in 2015\cite{Tabuchi2015}\,. For downscaling and integration, however, smaller YIG structures are needed. Taking these results into account it is a promising perspective to realize a new transfer mechanism by coupling magnons to mechanical oscillations in a nanomechanical resonator via magnetoelastic coupling. Obviously, YIG would be an ideal candidate for these resonators since YIG is the material with the lowest known Gilbert damping\cite{Kasuya1961} and it exhibits extremely long lifetimes for spin waves (magnons) in the $\mu$s regime. As a single crystalline garnet material with a Young's modulus of the same order of magnitude as that of silicon carbide it is expected to also provide low losses for mechanical waves (phonons) and may yield nanoresonators with high quality factors. Again in macroscopic YIG spheres in the sub-mm range the coupling of magnons to phonons has already been demonstrated\cite{Zhang2016}\,. However, up to now no method was known to shape three-dimensional nanostructures from monocrystalline YIG. Nanopatterning of thin films with reasonable quality has been demonstrated\cite{Jungfleisch2015, Li2016, Collet2016, Zhu2017, Collet2017}\,, but no patterning of nano-sized free standing resonators has been put forward.
Nevertheless, it would be extremely attractive if micron- or sub-micron sized YIG bridges or cantilevers were available. The mechanical resonance frequencies in such structures may be easily engineered to fall in the range of typical magnon frequencies\cite{Huang2003}\,. As a first step in this direction we have realized the fabrication of freely suspended YIG microbridges with very low damping for spin waves. Although the mechanical properties could not yet be investigated in detail, mechanical resonance frequencies calculated for their dimensions using the elastic properties of YIG fall into the range of several hundred MHz and may even reach the GHz regime.

\section{3D nano fabrication}

Fabrication techniques for suspended single crystal nanostructures mostly use subtractive processing by removing material from a single crystal (bulk or layer). The most straight forward method uses focused ion beam (FIB) lithography to directly shape the desired structure from bulk or thin film\cite{Babinec2011}\,. Although very flexible in terms of possible geometries this technique suffers from the possible damage to the crystal structure by extended beam tails which might be detrimental for the magnetic properties of YIG. Also it requires lateral access for the beam in order to remove the material underneath the suspended structure preventing the creation of multiple structures in close vicinity. Alternatively a crystalline film (resonator material) may be deposited on top of a sacrificial layer. The resonator itself is shaped by lithography and dry etching and only becomes free-standing when the underlying sacrificial layer is removed by highly selective wet chemical etching\cite{Schwarz2000, Carr1997}\,. The resulting geometry, however, has several limitations. It is not truly three dimensional but only a partly suspended two dimensional structure. Also the suspended resonator must be more narrow than the un-suspended pads to which it is attached. Otherwise the pads are under-etched during the removal of the sacrificial layer. Unfortunately no sacrificial layers are known for high quality crystalline YIG films which can only be deposited on garnet surfaces (especially gallium gadolinium garnet, GGG) and no selective wet etchants are available for these materials.

On the other hand nanoscale additive fabrication of polycristalline materials is achieved by electron beam lithography, evaporation, and lift-off. A typical example is the fabrication of metallic air bridges, well known since more than a decade \cite{Yacoby1995, Sherwin1994,Borzenko2004}\,. The process allows for densely packed structures with high flexibility in terms of geometry. However, it requires low temperature deposition of the material because of the limited thermal stability of electron beam resists. This prevents its use for the patterning of monocrystalline materials such as YIG, which in most cases need to be deposited at elevated temperatures.

A new kind of deposition method for thin film YIG has recently been demonstrated. Amorphous YIG films are deposited at room temperature on GGG using either pulsed laser deposition\cite{Hauser2016, Hauser2017} or sputtering\cite{Chang2014}\,. In a subsequent annealing step the material adapts to the lattice structure of the substrate resulting in thin single-crystalline YIG films. Surprisingly, the quality of these films in terms of damping surpasses the quality of thin films deposited at high temperature\cite{Hauser2016, Hauser2017, Chang2014}\,. Because deposition is done at room temperature this deposition method is compatible with electron beam lithography. In this way the fabrication of laterally nanopatterned YIG with reasonably small Gilbert damping constants has been demonstrated recently\cite{Jungfleisch2015, Li2016, Zhu2017}\,. Theoretically, this process also allows the fabrication of beams and bridges when it is adapted to the patterning process used for metal bridges described above. Nevertheless, the higher kinetic energies of the deposited particles in pulsed laser deposition compared to evaporation may necessitate a specially adapted resist profile to guarantee a successful lift-off. Further on the recrystallization is more challenging. In a thin film, crystallization needs to progress only vertically from the substrate to the film surface (with a typical distance of 100 nm or less). In a bridge structure, however, the crystallization starts at the base of the supporting pillars which are in contact with the substrate and then needs to progress around bends across the entire span of the bridge in order to achieve a monocrystalline structure. Any additional nucleation site for crystallization may disturb the process and introduce an additional grain boundary.
As we show in the following, it is possible to realize such a 3D lift-off process for YIG with the crystallization (which indeed starts at the substrate) extending throughout the complete bridge structure even over distances of several micrometers.

\section{Processing}
\begin{figure}
\includegraphics[width=0.8\textwidth]{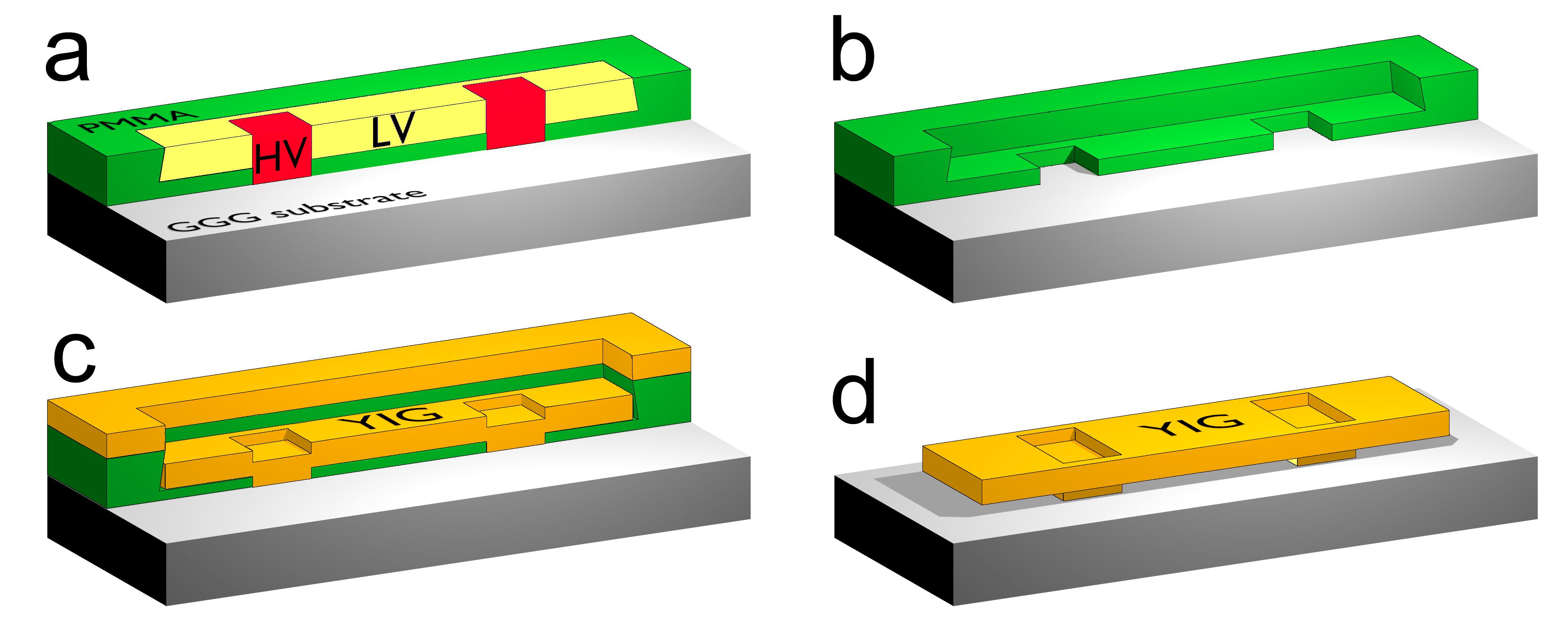}
\caption{Schematic drawing of the patterning process. (a) A resist (green) is exposed with two different acceleration voltages. A low voltage exposure is used for the span of the bridge (yellow) and a high acceleration voltage (red) exposes the pillars down to the substrate. (b) After development the void in the resist has the shape of the bridge and a slight undercut which later facilitates the lift-off. (c) The YIG is deposited and the shape of the bridge becomes visible. It is important that the YIG on the resist surface is well separated from the bridge itself. (d) After lift-off a free-standing bridge is obtained. }
\label{process}
\end{figure}

\begin{figure}
\includegraphics[width=0.6\textwidth]{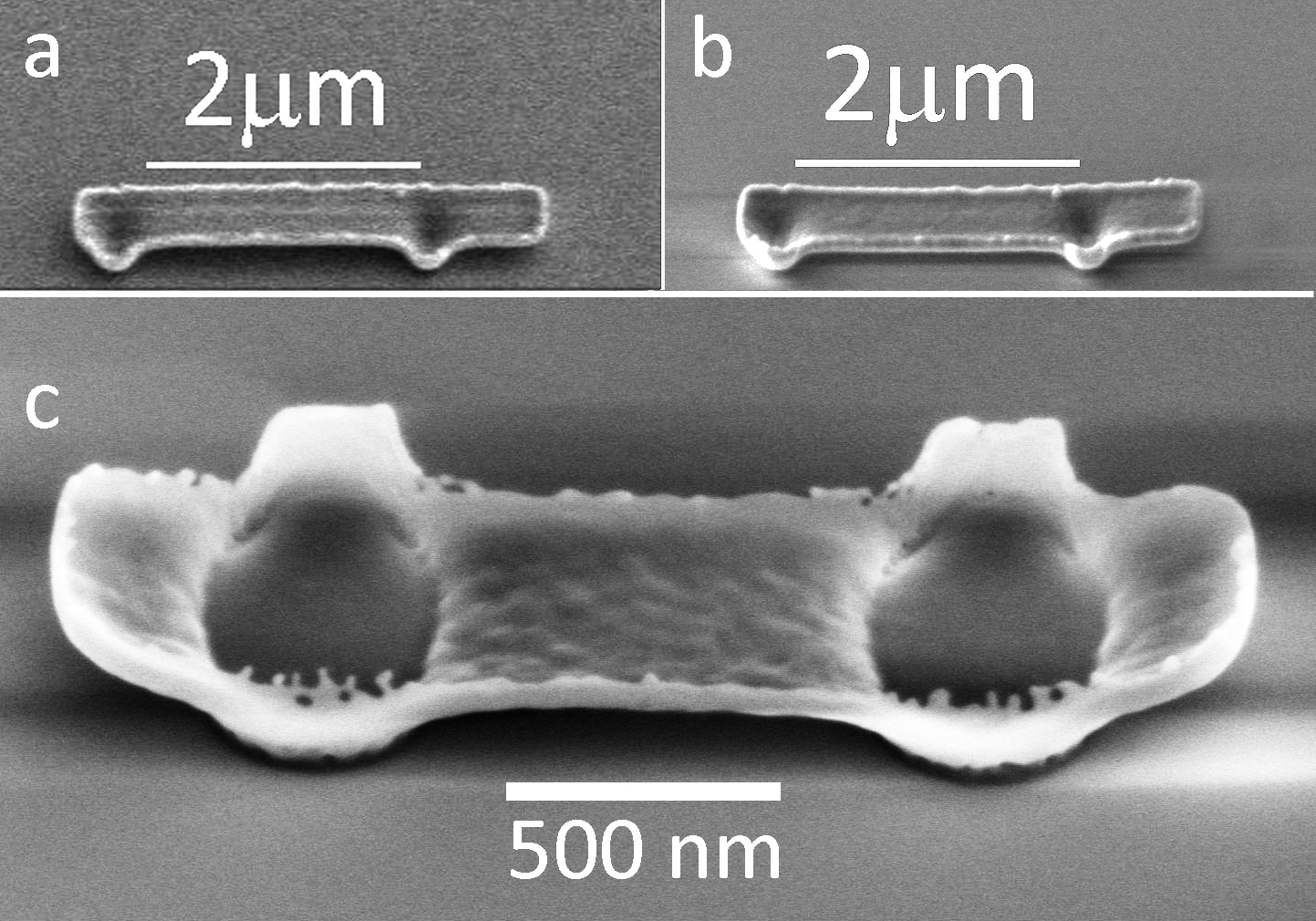}
\caption{SEM images of two different bridges. (a) and (b) show a larger bridge before and after annealing , respectively. (c) shows a smaller bridge after annealing. The deposited YIG has a nominal thickness of 110 nm.}
\label{SEM}
\end{figure}

Figs.\,1a-d schematically show the applied process flow. A thick PMMA layer on a \textless111\textgreater oriented GGG substrate is patterned using electron beam lithography at different electron acceleration voltages for the span (low voltage/LV) and pillars (high voltage/HV) of the bridges, respectively (Fig.\,\ref{process}a). Further details are provided in the methods section. The resulting structure after development of the e-beam resist is shown in Fig.\,\ref{process}b.  It exhibits holes down to the substrate for the pillars and a groove for the span of the bridge. At the sides the groove has a slight undercut which later facilitates the lift-off process. Onto the developed structure the amorphous YIG material is deposited by PLD at room temperature (Fig.\,\ref{process}c). Subsequent lift-off and resist removal results in a bridge structure (Fig.\,\ref{process}d) which is finally annealed.
Fig.\,\ref{SEM}a shows a scanning electron microscopy (SEM) image of a YIG bridge prior to (a) and after annealing (b). The bridge has a nominal span length of $\rm2\,\mu m$  and a YIG layer thickness of approximately $\rm 110\,nm$. The length of the span does not change during the annealing step within the measurement accuracy of the SEM.
For the experiment shown here the pillars are not placed at the end of the bridges. This design yields an overhang at the end to combine the investigation of short cantilevers fixed on one end only with that of bridge structures which are clamped at both ends. The resulting bridges and cantilevers are flat and strain free after the lift-off. Subsequent to annealing the bridge itself remains mostly unchanged, however, the overhang is bent upward (Fig.\,\ref{SEM}b) indicating the presence of strain.

During the crystallization at more than than 800$^\circ$C the lattice can reorder and a structure with very little or no strain is created. During cool-down, however, the difference in thermal expansion coefficient of YIG and GGG can lead to a small deformation. The YIG now exhibits tensile strain. While in a continuous layer on a substrate this strain would lead to a change in lattice constant the bridge can now follow the strain by deformation. By tilting the feet inward, the length of the span can be decreased while the tilting can of the feet can lead to the small upward bend of the overhang. The thermal expansion coefficients for YIG is smaller than that of GGG by $\rm\sim 2\times10^{-6}\,K^{-1}$. By cooling from 800 $^{\circ}$C to room temperature the contraction of the YIG lattice would be approximately 0.1\% larger than for GGG. It should be noted that any resulting shortening of the bridge is too small to be measured with the accuracy of our electron microscope.

Fig.\,\ref{SEM}c shows a close up view of an annealed YIG bridge with a span of 750 nm also after annealing. The deposited YIG has a nominal thickness of 110 nm. The edges of this bridge are quite rough and show a lot of residue from the lift-off process. Obviously, these can be detrimental for the quality of mechanical resonances. As we show later, these residues can mostly be avoided or removed.

\section{Structural characterization}
\begin{figure}
\includegraphics[width=0.8\textwidth]{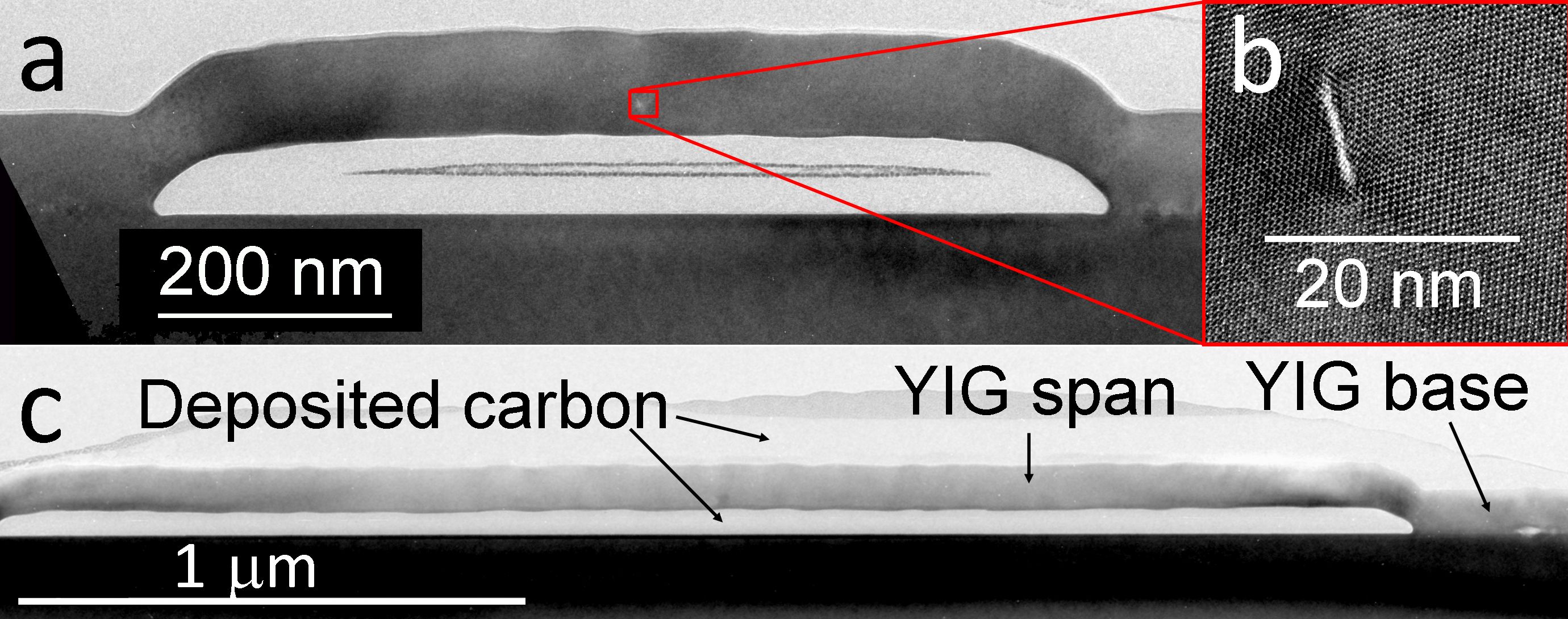}
\caption{Transmission electron micrographs for bridges with a nominal thickness of the span of 110 nm. (a) Shows a bridge with a span of approximately 850 nm length and a height of 75 nm underneath the span. (b) Higher magnification shows single crystalline material with a single defect in the center of the bridge. (c)  TEM cross section of a bridge with increased length. Even for a length of 2.8 $\mu$m the bridge is free of defects except for the central defect. Above and below the bridge a carbon film is visible which has been deposited using the electron beam during TEM preparation to protect the surface of the bridge.}
\label{TEM}
\end{figure}

While the SEM images show that the molding of the material is successful, the local crystalline quality can only be assessed by transmission electron microscopy (TEM). Atomic resolution TEM has been performed on different bridges after annealing (details described in the methods section). Fig.\,\ref{TEM}a shows a cross-sectional view of a small bridge with a span of approximately 850\,nm and a height between span and substrate of 75\,nm. The sample was prepared using a focused ion beam and cut along a \{011\} plane perpendicular to the surface. The viewing direction of the TEM is along \textless011\textgreater with a small tilt angle.

The pillars which are in direct contact with the substrate show an epitaxial monocrystalline lattice as also observed for large area deposition by Hauser \textit{et al.} \cite{Hauser2016}\,. The transition to the span where the material is thinner shows a number of defects likely due to partially relieved shear strain that can be expected in this location.

\begin{figure}
\includegraphics[width=0.9\textwidth]{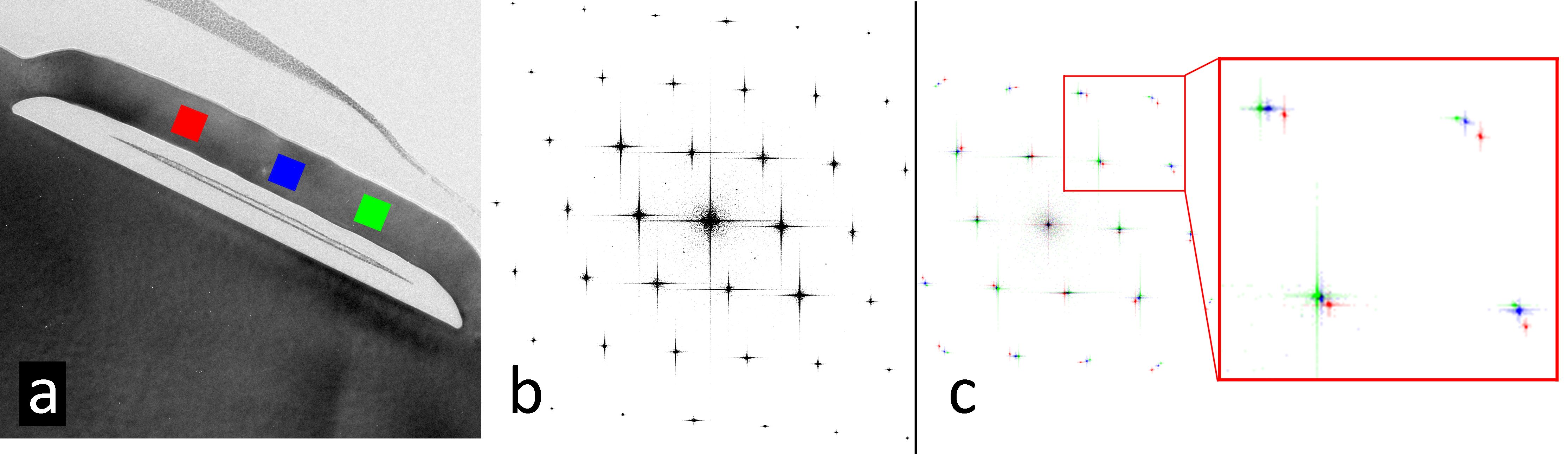}
\caption{Fast Fourier transforms of different parts of the lattice of a single bridge. (a) shows a TEM image of a bridge with three different square areas color coded in red, green, and blue. (b) shows an FFT of the substrate. (c) For the color coded areas the FFT of the lattice is superimposed using the same color code. A zoom into the superposition (frame) shows that a very small rotation of the lattice has taken place which is in the range of $\sim 1^\circ$. All FFTs are obtained from images with the same orientation and magnification.}
\label{FFT}
\end{figure}

The span of the bridge, however, appears monocrystalline and of perfect crystallinity except for a single defect in the center (Fig.\,\ref{TEM}b). This defect is a consequence of the crystallization process as described below. To investigate possible differences in lattice orientation of substrate and bridge FFts of TEM images were taken at different spots of the sample. A comparison of FFTs from the substrate and the bridge shows that except for a minute lattice rotation the lattice parameter and orientation are identical for substrate and bridge. This is expected due to the excellent lattice match between YIG and GGG. (mismatch $\sim 0.06\%$). In addition FFTs from different points of the bridge are superimposed to see whether the lattice orientation varies along the bridge (Fig.\,\ref{FFT}). A color coded overlay of the FFTs on left, right, and center of the span shows that the crystal orientations on both sides are tilted with respect to each other with a tilt angle of about $1\,^\circ$. A similarly small rotation is observed when comparing FFTs from bridge and substrate. From these results we can deduce that crystallization starts simultaneously at both pillars, where the material is strained. Thus the two crystallization fronts may be slightly tilted with respect to each other. When they meet at the center of the span the resulting mismatch can only be compensated for by the formation of the crystal defect such as a small angle grain boundary observed at the center of the bridge. In addition, this mechanism explains the small rotation of left and right hand part of the bridge with respect to each other and with respect to the substrate.

To investigate the influence of the bridge size on crystallinity also cross-sectional TEM images of longer bridges are studied (Fig.\,\ref{TEM}c). Even for a length of $2.8\,\rm\mu$m a similar quality of the span (which is the functional part of the resonator) is obtained.

\section{Spin dynamics}

Because of the reduced amount of material it is not possible to measure the saturation magnetization M$_{\rm S}$ of the bridges directly with magnetometry methods. From previous experiments we know that YIG layers fabricated by room temperature deposition and annealing under similar conditions exhibit M$_{\rm S}$ up to 27\,\% below the bulk value of $\mu_0 M_{\rm S}\approx 180\,mT$\cite{Hansen1974}\,. We would like to note that the M$_{\rm S}$-value used for the micromagnetic simulations (132\,mT) is in excellent agreement with these results.

\begin{figure}
\includegraphics[width=0.8\textwidth]{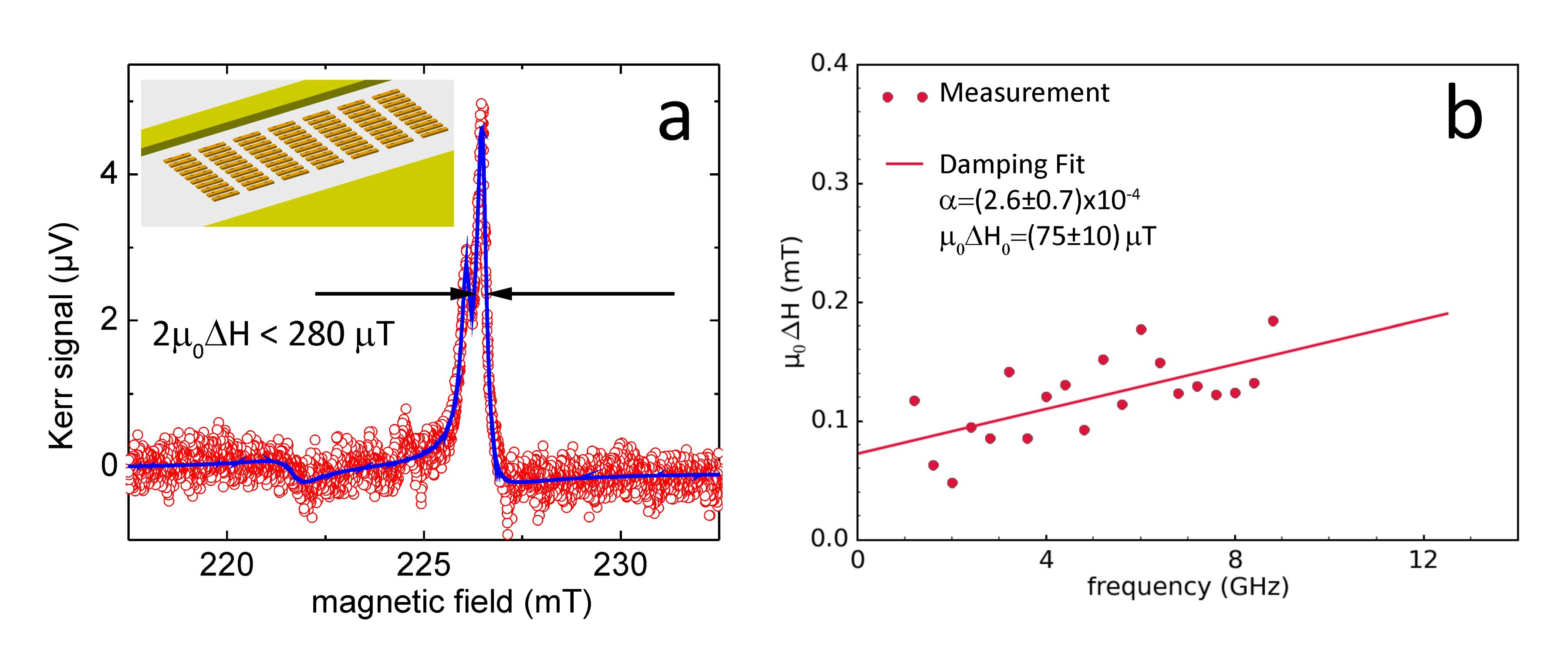}
\caption{(a) FMR spectrum obtained by TR-MOKE at the center of a single bridge, excited at 8 GHz. The red circles show the measured data while the blue line is a fit using three Lorentzian line shapes. The arrows are a guide to the eye showing an upper limit for the full width at half maximum which is 2$\mu_0\Delta$H. The half width at half maximum $\mu_0\Delta$H is mostly referred to in literature as the line width. The measurement which is performed on a single spot with a diameter of approx. 300\,nm shows two very sharp lines with a small overlap. The line width $\mu_0\Delta$H is smaller than 140\,$\mu$T. The insert shows a sketch of an array of bridges located between signal line and ground of a CPW. (b) Line width plotted versus frequency. A least mean square fit yields a slope corresponding to a Gilbert damping of (2.6$\pm$0.7)$^{-4}$.}
\label{LINEDAMPING}
\end{figure}

In order to obtain a detailed and accurate measurement of the local dynamic properties we perform time-resolved scanning Kerr microscopy (TR-MOKE) experiments on a 110 nm thick YIG-bridge. Using this  method it is possible to image directly the different resonant magnon modes in individual bridge structures. To achieve the necessary high frequency excitation of the YIG structures an impedance matched coplanar wavegude (CPW) is deposited by electron beam lithoghraphy and lift-off processes onto the sample. The CPW is positioned such that an array of bridges is located in the gap between signal line and ground plane (inset of Fig.\,\ref{LINEDAMPING}a). The investigated bridge has a width of 600\,nm and a span length of $ 3\,\rm\mu m$. The thickness of the deposited YIG film is 110\,nm and the gap under the span is 100\,nm. The sample was deposited using the parameters described in the methods section.

\begin{figure}
\includegraphics[width=0.8\textwidth]{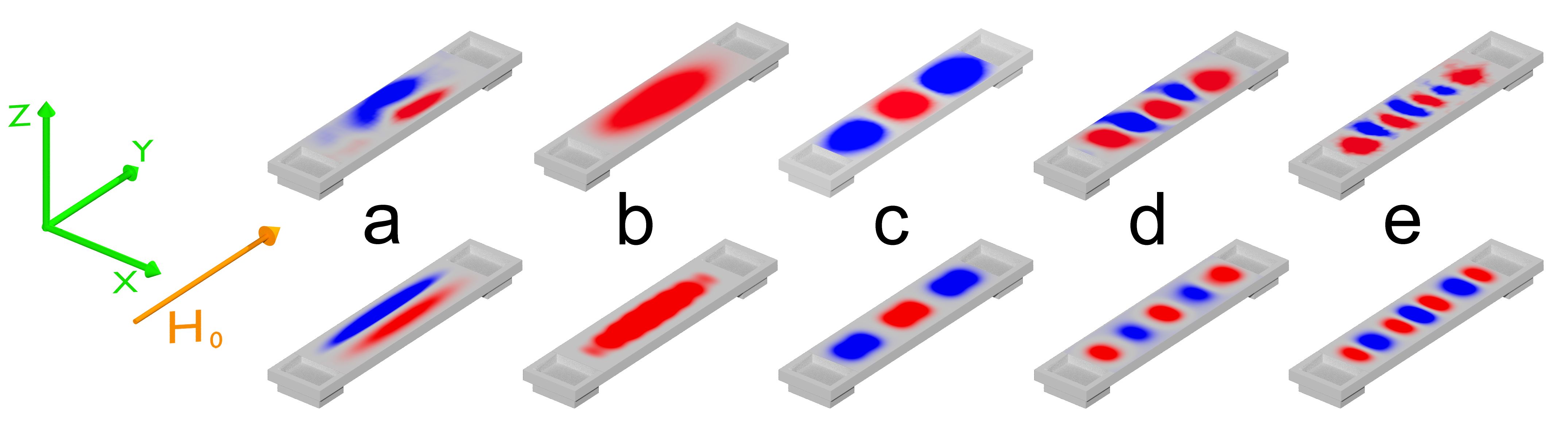}
\caption{Time-resolved scanning Kerr microscopy (TR-MOKE) images of standing spin-wave modes and simulations. The top row shows TR-MOKE results for the main mode (b), one Damon Eshbach mode (a), and three different backward volume modes (c-e). Measurement parameters are (magnetic field/excitation frequency) 11.96\,mT/2\,GHz (a), 21.95\,mT/2\,GHz (b), 25.61\,mT/2\,GHz (c), 89.72\,mT/4\,GHz (d), and 92.52\,mT/4\,GHz (e). The modes were imaged at the peak amplitude of the respective resonance. The bottom row shows the corresponding simulation results from simulations at fixed respective magnetic fields (see also methods section). Simulation parameters are 19.4\,mT/2.32\,GHz (a), 19.4\,mT/2.00\,GHz (b), 19.4\,mT/1.85\,GHz (c), 83.8\,mT/3.73\,GHz (d), and 83.8\,mT/3.66\,GHz (e). The coordinate system on the left hand side shows the orientation of the external magnetic field H$_0$.}
\label{MOKE}
\end{figure}

The spatially resolved measurements are performed with the external magnetic field oriented along the bridge allowing for the excitation of the backward volume modes (BVM) with k-vectors along the bridge and the Damon Eshbach modes (DEM) with k-vectors at an angle of 90$^\circ$. Fig.\,\ref{MOKE} (top row) shows a number of different modes for increasing magnetic field. The fundamental mode with only one antinode is shown in Fig.\,\ref{MOKE}b. Three standing BVM with nodes distributed along the bridge are shown in Fig.\,\ref{MOKE}c-e, while a DEM mode shows a node extending along the bridge (Fig.\,\ref{MOKE}a). It is clearly visible that the magnons are localized in the span of the bridge and no direct coupling to the pillars or beyond is observed.

\begin{figure}
\includegraphics[width=0.6\textwidth]{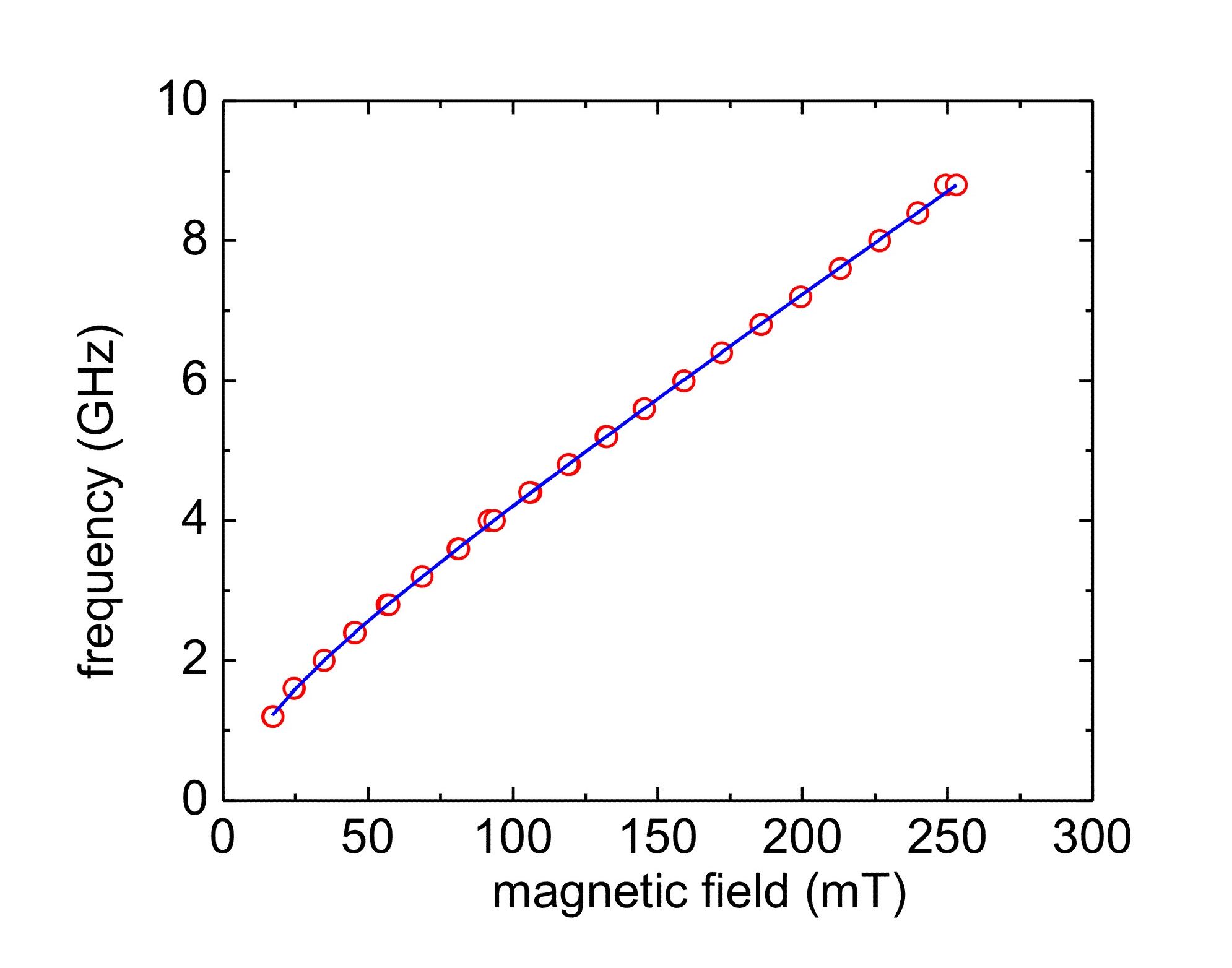}
\caption{Resonance frequency plotted as a function of applied magnetic field. The results nicely agree except for small deviations at low magnetic fields. The red circles show the measured data while the blue line is the respective fit using the Kittel formula.}
\label{Kittel}
\end{figure}

We have also modelled the different magnon modes using MuMax3\cite{Vansteenkiste2014}\,. Fig.\,\ref{MOKE}$\\$(bottom row) shows the respective simulations, which are in good agreement with our experiments. Like the bridge investigated by TRMOKE the simulated bridge has a width of 600\,nm and a span length of $ 3\,\rm\mu m$. The thickness of the deposited YIG film is 110\,nm and the gap under the span is 100\,nm. The gyromagnetic ratio $\gamma$ obtained in the simulations is 178\,GHz/T which is close to the value obtained from the MOKE data (171\,GHz/T, Fig.\,\ref{Kittel}). The saturation magnetization was fitted to match the spin wave patterns resulting in a value of $\rm\mu_0 M_{S}\approx 132\,mT$ which is in good agreement with that of large area films deposited by the same method\cite{Hauser2016}\,.

In order to obtain a better understanding in terms of the magnetization dynamics in the YIG nano bridges FMR spectra are measured by TRMOKE on a single spot in the center of the bridge for several frequencies. Such a resonance spectrum is shown in Fig.\,\ref{LINEDAMPING}a where the main resonance peak has a line width of  approximately 140 $\rm\mu$T at 8 GHz. This value is among the smallest values reported for PLD grown thin film material so far. Only material grown by liquid phase epitaxy exhibits smaller linewidths. From our data we find a Gilbert damping value of the main resonance of $\alpha\approx$(2.6$\pm$0.7)$\times$10$^{-4}$ (Fig.\,\ref{LINEDAMPING}b). Also this value is lower than all values reported for YIG grown by PLD at elevated temperatures. The inhomogeneous line width at zero field is $\mu_0\Delta H_0=75\pm10\rm\mu$T which is lower than anything reported for PLD grown thin film YIG so far. For the given configuration these numbers can also be translated into spin wave life times resulting in 220 ns (3.2 GHz), 160 ns (5.2 GHz), and 120 ns (8.4 GHz).

We also determine the effective saturation magnetization M$\rm_{eff}$ which also contains any anisotropy and the gyromagnetic ratio $\gamma$ the resonance fields of the main FMR line are determined as a function of frequency (Fig. \ref{Kittel}) and the data is fitted by the Kittel formula:

\begin{equation}\label{KE1}
\omega =\mu_{0}\gamma\sqrt{H\rm_{{FMR}}(H\rm_{{FMR}}+M\rm_{eff})}
\end{equation}

The fit yields $\gamma$=(180.3$\pm$0.6)\,GHz/T and $\rm\mu_0 M_{eff}$=(0.125$\pm$0.003)\,T.

\begin{figure}
\includegraphics[width=0.6\textwidth]{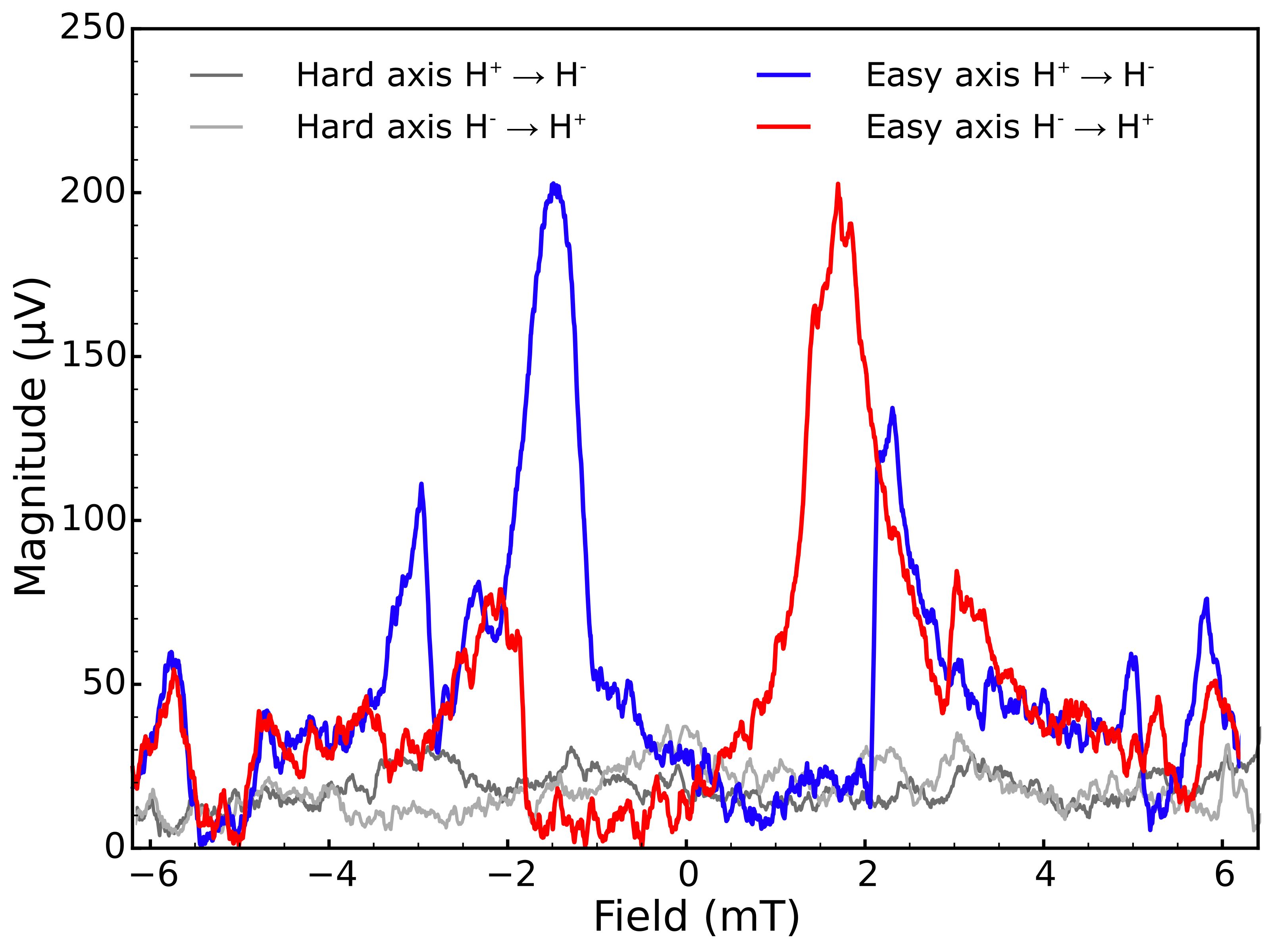}
\caption{TRMOKE measurement of the static switching behavior. While sweeping the field through the static hysteresis the magnitude of the rf-suceptibility is determined as a function of the applied magnetic field. The hysteretic part of the measurement represents the hysteresis of the static switching of the magnetization.}
\label{Switching}
\end{figure}

In addition to the dynamic properties TR-MIKE also allows to investigate the static switching behavior of individual nanobridges. For this we use the method in an off-resonant fashion around zero field (Fig.\,\ref{Switching}). Here the phase and the magnitude of the rf-susceptibility are used to detect the switching as first demonstrated in \cite{Woltersdorf2007b}\,. For the measurement the microwave frequency is set to 1 GHz. The probing light spot is placed at the center of the same bridge. The magnitude of the response depends on the internal magnetic field and is therefore sensitive to the relative alignment of magnetization and applied magnetic field. Hysteretic behavior is found when magnetization and applied magnetic field are antiparallel. From this we determine a coercive field of $\mu_0 H_{\rm C}\approx2\,\rm mT$ for the bridge (lateral dimensions of the span: 600\,nm$\times$ 3\,$\mu$m) when the magnetic field is aligned with the long axis of the bridge structure (easy axis). For the magnetic field aligned along the short axis (hard axis) of the bridge we find no hysteretic behavior as expected. This coercive field is considerably larger than for comparable continuous YIG films of the same thickness where we find coercive fields of less than 0.1 mT \cite{Hauser2016}\,. The enhanced coercive fields in the YIG nano bridges are expected and a consequence of the shape anisotropy and the increased contribution of the domain wall nucleation energy to the magnetization reversal in nanostructures.


In the FMR spectrum also a second line is visible which partly overlaps with the main peak. Spatially resolved measurements indicate that the two halves of the span which are separated by the central crystalline defect differ in resonance field by approx. 100$\rm\mu$T at 8 GHz. This can be explained by the rotation of the two sides observed in transmission electron microscopy. When the field is applied exactly along one half of the bridge, the small tilt of the other half can shift the resonance field in the order of 100\,$\rm\mu$T at 8 GHz simply because a very small demagnetizing field is added to the external field. For one degree of tilt this modification can be as large as 0.05\% of the resonance field which is sufficient to explain the observed resonance line shift.

\begin{figure}
\includegraphics[width=0.8\textwidth]{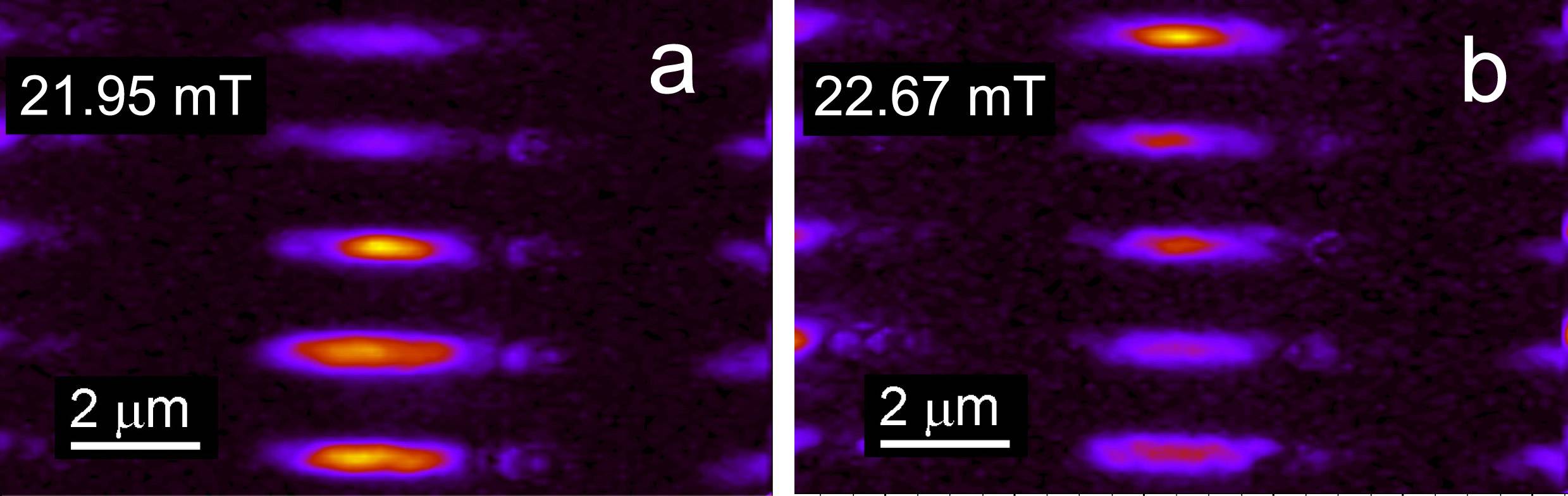}
\caption{Two TR-MOKE images obtained at a frequency of 2 GHz showing five adjacent bridges at two different magnetic fields, respectively. In both images at least one of the bridges shows an intense resonance of the mode with one antinode only. Apparently the resonance field between bridges can vary at least by 0.8\,mT.}
\label{Variation}
\end{figure}

In addition the spatial resolution of the TR-MOKE also allows to investigate the variation of the resonance field between different bridges and between different parts of a single bridge (namely span and overhang), respectively. Fig.\,\ref{Variation} shows TR-MOKE images of five different bridges obtained simultaneously and repeated for two different magnetic fields but at the same excitation frequency.  For individual bridges the main resonance (only one antinode) appears at fields that vary by almost 0.8\,mT, respectively.

\begin{figure}
\includegraphics[width=0.4\textwidth]{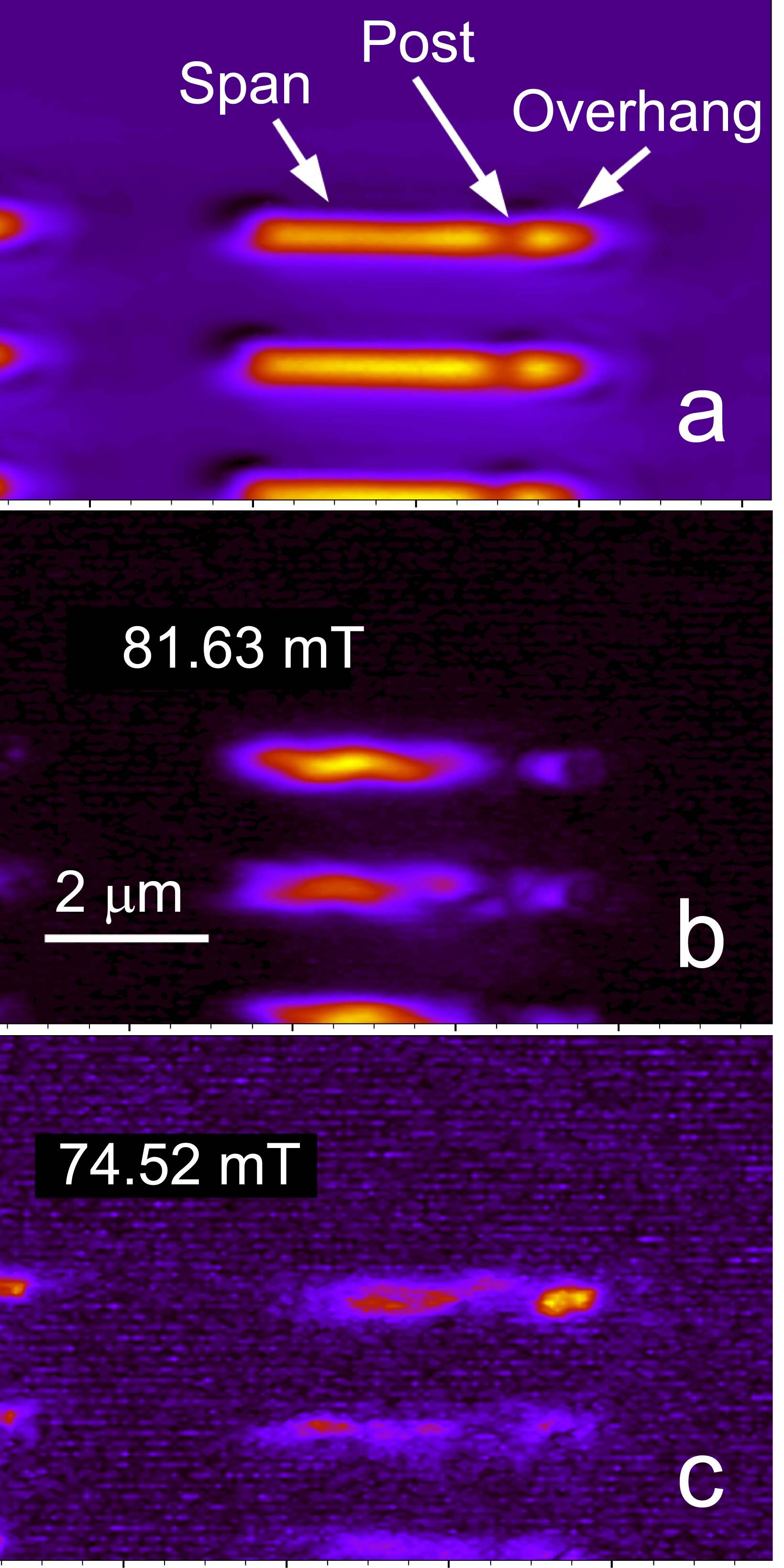}
\caption{Optical topography image of several bridges (a) and two TRMOKE images of the same area acquired at a frequency of 6 GHz at different respective magnetic fields (b-c). In the topography image we can clearly discern the span of the bridge, the base which is slightly darker and the overhang at the end. (b) Shows the main resonance of the span with one antinode while in (c) a similar mode for the overhang of the same bridge is observed. The resonance fields differ by 7\,mT.}
\label{End}
\end{figure}

The resonance in the overhang of a bridge can only be imaged by sweeping the field over a wider range.  Fig.\ref{End} shows that the overhang also exhibits a localized resonance. The resonance field, however, is offset by approx. 7 mT from the main resonance field of the corresponding span. This shift can be caused by the different strain in the span which is pinned on both sides and the overhang which is pinned only on one side as well as by the different size of the two regions. As the resonance does not extend into the foot of the bridge the k-vector is determined by the length of the area on resonance as we no longer observe a true uniform mode but a standing spin wave with zero nodes.

\begin{figure}
\includegraphics[width=0.8\textwidth]{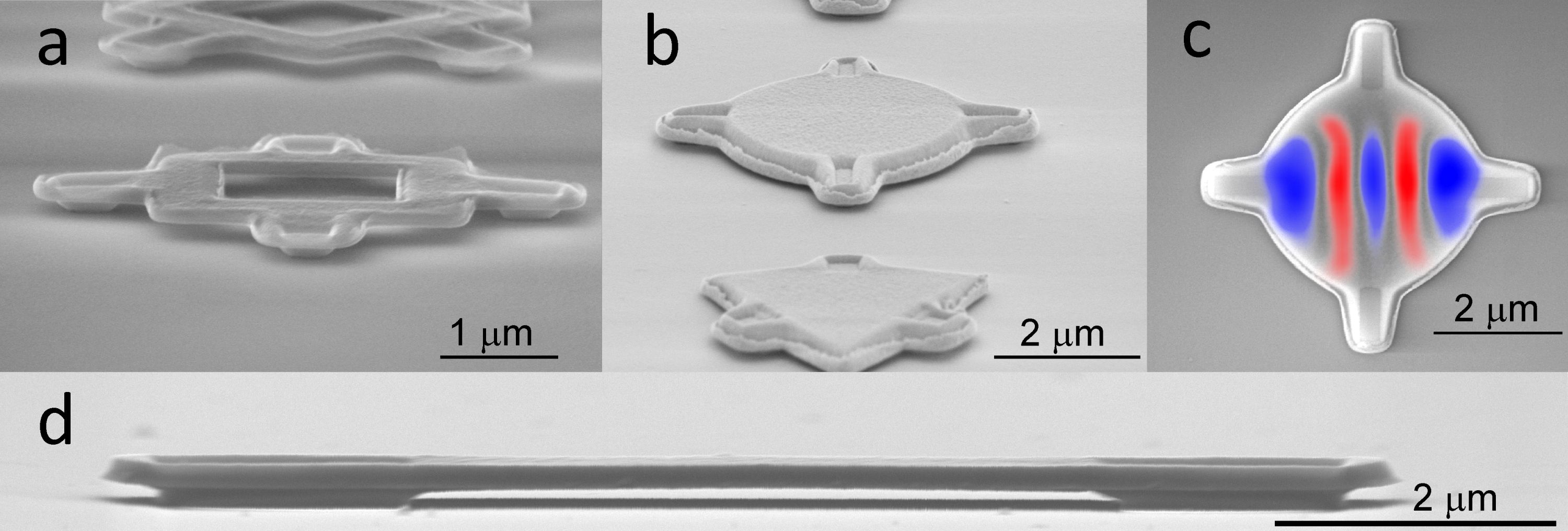}
\caption{SEM images of more complex resonators. The process allows to fabricate various shapes such as open squares (a) or disks and triangles (b). (c) TR-MOKE image of a standing backward volume mode measured on a disk resonator overlayed to an SEM image of the same structure. For these structures the nominal YIG thickness is 210 nm. (d) shows a bridge on which a post-annealing wet-etch was applied. The artifacts at the seam of the structure are strongly reduced.}
\label{Outlook}
\end{figure}

It should be noted that the fabrication process is not limited to simple bridge geometries but highly flexible and can be extended to more complex structures as shown in the examples of Fig.\,\ref{Outlook} paving the way to a number of applications and experiments. Again, also the magnetic excitations are well defined and can be directly imaged. SEM image and MOKE data in Fig.\,\ref{Outlook}c are obtained from the very same structure. We have also tried to reduce edge and surface roughness which may deteriorate the mechanical resonance properties by using an optimized multi-layer resist and a post-annealing wet-etch step. As a result an improved bridge with smoother edges is shown in Fig.\,\ref{Outlook}d.

\section{Discussion}
It is possible to fabricate 3D YIG nanobridges using electron beam lithography, room temperature PLD and lift-off. The structural characterization shows that crystallization during the annealing process progresses throughout the bridge on a length scale of more than one $\mu$m leading to an undisturbed lattice with only very few defects. The span of the bridge typically contains a single crystal defect. To the best of our knowledge, until now this kind of long range crystallization process throughout a 3D nanostructure has not been reported. The damping does not reach the record values of low temperature grown YIG layers but is still in the range of high quality PLD grown YIG films. The minimum line width of 140\,$\mu$T at 8 GHz for a single bridge is well in the range of high quality thin film material and various resonant magnon modes can be identified in scanning TR-MOKE. Both line width and damping thus rival those obtained for large area thin films deposited at higher temperatures. The mechanical resonances of the YIG bridges have yet to be characterized, nevertheless, an estimate of possible resonance frequencies can be given. According to Yang \textit{et al.}\cite{Yang2001} the resonance frequency of a so called doubly clamped beam which corresponds to the span of our bridge is approximately

\begin{equation}\label{KE1}
f_{res}\approx1.03 \frac{t}{L^2}\sqrt{\frac{E}{\rho}}
\end{equation}

with E the Young's modulus of YIG ($2\times10^{11}$\rm Pa)\cite{Clark1961}\,, $\rho$ the density (5.17 g/cm$^2$)\cite{Clark1961}\,, $t$ the thickness and $L$ the length of the beam. Using these parameters with a thickness  of 150 nm and a length of 1 $\mu$m a resonance frequency of 964 MHz is expected while the same beam with a length of 500 nm resonates at 3.86 GHz, which is well in the range of typical magnons as measured in our experiments. For further development of the method the next steps will be to fabricate more complex resonators that can not only host magnons with high quality factors but are also suitable for the characterization of mechanical vibrational modes. In addition statistics need to be obtained by TR-MOKE on the variation and reproducibility of resonance frequencies in nominally identical resonators, which are crucial for applications where the exact behavior needs to be predictable.

Possible applications for the nanoresonators can be found in various areas. Spin cavitronics for example investigates strong coupling of cavity resonator modes to magnon modes in macroscopic magnetic samples (typically YIG). In these experiments current technology uses large volume YIG samples coupled to macroscopic planar superconducting microwave resonators \cite{Huebl2013} or large cavities \cite{Tabuchi2014}\,. Our YIG nano resonators might be deposited over micron sized superconducting coplanar waveguides allowing for more complex experiments. In spin caloritronics YIG bridges may be used to create large and extremely and well defined temperature gradients because the span of the bridge is thermally decoupled from the substrate. It should be noted that a temperature difference of only 1 K over a bridge with a length of 1\,$\mu$m corresponds to a temperature gradient of 10$^6$ \,K/m. If coupling between phonons and magnons in the nanoresonators can be established even an application for readout of qubits or conversion of quantum information between the microwave and the optical regime may be possible. Therefore the new technology platform presented here may pave the way for downscaling allowing these schemes to be realized on the micron scale or below and facilitating future integration of qubits.

\section{Methods}

\subsection{Electron beam lithography}
The pillars and the span of the bridge are exposed using PMMA as a resist and two different respective acceleration voltages. The span is exposed at 2.8\,kV while the acceleration voltage for the span is 4.5\,kV. For both exposures the area dose is 100\,$\mu$C/cm$^{2}$. The structures are developed for 60 s in isopropanol.

\subsection{Pulsed laser deposition of YIG}
The YIG is deposited in 0.025\,mbar of oxygen from a home-made target. Laser parameters are 248\,nm wavelength,  fluence of 2.5\,J\,cm$^{-2}$, and a repetition rate of 5\,Hz. Annealing is performed in an oxygen atmosphere (99.997\%) at ambient pressure and 800$^\circ$C for 3\,hours.

\subsection{TEM preparation}
TEM samples from bridges are prepared using a focused gallium ion beam ‘FEI VERSA 3D’ dual beam microscope by the classical FIB in-situ lift-out technique as described for instance by Bals \textit{et al.}\cite{Bals2007}\,. Due to the electrically isolating substrate this procedure is extended for the preparation of the sample after thermal treatment by depositing a thin conductive carbon layer via ion sputtering before transferring the sample to the FIB. As the first step in the preparation procedure inside the FIB a 200\,nm thick carbon layer is deposited locally using the electron beam at 5\,kV from the top through the bridge to fill the space under the bridge with carbon. The hole under the bridge is filled by locally cracking the organometallic complex gas from the platinum Gas Injection System of the FIB with a 5\,kV electron beam. After lift-out the TEM lamellae are mounted to a grid, thinned down to a thickness below 150\,nm, and stepwise cleaned on both sides from amorphous material by operating the ion beam of the FIB at 5\,kV, 2\,kV and 1\,kV. HRTEM images from these samples are obtained using a JEOL JEM-4010 TEM operated at 400\,kV.

\subsection{TR-MOKE}
For the time resolved magneto optic Kerr (TR-MOKE) measurements we use a frequency doubled fs-laser operating at 520 nm to illuminate the sample in a scanning optical microscope with polarization analysis. A detailed description of this method is presented in the work of Farle {et al.}\cite{Farle2013}\,. In our TR-MOKE measurements the magnetization is excited by continuous wave microwave magnetic field which is phase synchronized to the optical probe pulses, i.e. the sampling is stroboscopic. In order to allow for lock-in amplification of the magneto-optical signal the rf-excitation is modulated\cite{Woltersdorf2007, Stigloher2016}\,. The spatial resolution of the measurements presented in this manuscript is diffraction-limited to about 300 nm.

\subsection{Micromagnetic simulation}
The simulations were carried out using MuMax3. The simulated structure is a bridge with a rectangular span of 2700\,nm x 600\,nm x 110\,nm (l x w x h). The pillars are 300\,nm x 600\,nm x 110\,nm. After relaxing magnetization of the structure in the external magnetic field H$_0$ in x-direction a small field step perpendicular to the surface (z) is applied. The following precession in the z/y-plane is recorded and a FFT is performed. In the FFT the main oscillations are identified as peaks in the amplitude. The images are obtained by locally evaluating the amplitude and phase of the precession and transforming them into a color (intensity = amplitude, positive phase: red, negative phase: blue).

\subsection{Acknowledgment}

This work was supported by the German research foundation (DFG) via collaborative research centers SFB\,762 (TP B9) and TRR\,227 (TP B02). We thank the Max-Planck-Institut f\"{u}r Mikrostrukturphysik for making the JEOL JEM-4010 TEM available for our experiments.

\bibliography{sampleAPS}

\clearpage

\clearpage

\clearpage

\clearpage

\clearpage

\clearpage

\clearpage

\clearpage

\clearpage

\clearpage

\clearpage

\end{document}